# Subwavelength periodic plasma structures formed during the laser-pulse-induced breakdown within the transparent dielectric

V.B. Gildenburg [1,2,*], I.A. Pavlichenko [1,2]

[1] University of Nizhny Novgorod, Nizhny Novgorod 603950, Russia
[2] Institute of Applied Physics, Russian Academy of Sciences, Nizhny Novgorod 603950, Russia
[*] gil@appl.sci-nnov.ru

The spatiotemporal evolution of the field and plasma in the optical breakdown induced in the volume of transparent dielectric (fused silica) by the focused fs laser pulse is studied under condition of the so-called plasma-resonance-induced ionization instability that results in the deep small-scale periodic modulation of the breakdown plasma parameters in the direction of the laser polarization. In the framework of the model used, the optical electric field was calculated with allowance for the effects influencing both its long-scale structure (the beam focusing accounted for in the given-ray-tube approximation, phase and group delays, and back reflection) and the small-scale one (quasi-static enhancement in the plasma resonance regions). The plasma density evolution is described by the rate equation taking into account the photoionization, avalanche ionization, and ambipolar diffusion. Based on the fulfilled numerical calculations, we have described the main types of the breakdown wave originating in the focal region and have found the laser pulse intensity range where the instability evolving from very small seed perturbations leads to the formation of the contrast subwavelength periodic structure containing the number of the narrow zones with overcritical plasma density and enhanced energy deposition. The latter allows us to consider this structure as underlying the nanograting formation observed experimentally in the fused silica irradiated by series of repeated fs pulses.

As is well known, the so-called ionization-field instabilities developing during the breakdown of the medium by high intensity electromagnetic radiation of various frequency bands can lead to formation of small-scale periodic plasma structures instead the homogeneous plasma seemingly corresponding to the initially homogeneous radiation field [1-6]. During the last decade, the interest in such structures has been renewed in connection with their growing applications for the production of the subwavelength nanogratings in the optical materials, in particular in the fused silica, by the series of repeated laser pulses [7-24]. The creation of such nanogratings is considered now as a promising technique of the material optical properties modification and compact writing and storage of the optical information. This technique is nowadays widely demanded and already finds numerous applications in various functional devices based on three-dimensional photonic structures in the bulk of optical materials. Nevertheless the physical mechanisms predetermining these structures appearance are not revealed completely and remain the subject of discussion. As the likely initial cause underlying the nanograting formation in the transparent dielectric there were analyzed two different ionization-related mechanisms. The first of them is the breakdown around the randomly placed bubbles or atomic structure defects followed, according to numerical simulation results, by the progressive aligning the stochastic sites of ionization (supposedly due to the scattered waves interference [9,10,16,19-21] or the interface plasma wave excitation [17,18]) in more or less ordered rows. The second one (considered in Refs. [22,23] and being the subject of this paper) is the plasma-resonance-induced (PRI) ionization instability [1,3,5] having the quasi-static nature and developing from very small and slight seed perturbations of the plasma density and electric field due to their mutual enhancing effect against the homogeneous background. Its growth rate at the linear stage is maximal for 1D subwavelength modulation of discharge parameters in the direction of laser pulse polarization. An essential question determining the efficiency and practical importance of the latter mechanism is finding the conditions under which the spatially periodic plasma and field structures formed in a single fs laser pulse at the nonlinear stage of the PRI instability are of sufficient contrast to provide the formation of the contrast and deeply modulated grating in the material exposed to the series of the repeated laser pulses. It is this question that we address here, disregarding in particular from the important questions concerning the exact value of the formed structure period, which is some minor fraction (yet not defined exactly in both above approaches) of the wavelength in the ionized medium and will be considered as a given constant.

Both the linear and nonlinear stages of the transverse PRI instability in focused laser beams have been studied previously [1,3,5,22,23] only in the context of the 1D models devoid of any longitudinal (beam-parallel) structure caused by the beam focusing and pulse delay. In this work, we have studied development of the instability against the non-stationary and inhomogeneous background with allowance for both its transverse (small-scale) and longitudinal (long-scale) inhomogeneities, time delay effects, and the finite velocity of the breakdown wave propagating from the focal region in the direction opposite to the incident radiation. We analyze these effects based on the simplified approach using the separate description of the small-scale (transverse) and long-scale (longitudinal) structures formed in the near-axial region of the paraxial focused beam. The first of these structures is studied in the quasi-static approximation; for the description of the second one, we use the model of the focused wave beam with the given ray tubes.

In the framework of the model considered, the transverse electric field component of the $x$-polarized and $z$-propagated paraxial wave beam, with allowance for the transverse small-scale modulation of interest, can be written as $E_x = \text{Re}\{E(x,z,t)\exp(-i\omega t)\}$, where $\omega$ is the field frequency, $E(x,z,t)$ is its slow time-varying envelope. Its subwavelength ($x$-dependent) periodic structure is described by the equation

$$\frac{2i}{\omega}\frac{\partial E}{\partial t} + \varepsilon E = -\frac{c^2}{\omega^2}\frac{\partial^2 E_a}{\partial z^2}, \qquad (1)$$

where $\varepsilon = \varepsilon_m - (N/N_c)(1-i\nu/\omega)^{-1}$ is the permittivity of the ionized medium, $\varepsilon_m$ is its value in the absence of ionization, $\nu$ is the electron collision frequency, $N(x,z,t)$ is the plasma density, $N_c = m(\omega^2 + \nu^2)/(4\pi e^2)$ is its critical value, $E_a(z,t) = (1/\Lambda)\int_0^\Lambda E(x,z,t)\,dx$ is the $x$-averaged field, $\Lambda$ is the period of the small-scale structure of interest. Eq. (1) is in fact the constitutive equation; its right-hand side is the electric displacement being $x$-independent in the quasi-static approximation. The medium permittivity $\varepsilon_m$ is assumed to be initially slightly modulated: $\varepsilon_m = \varepsilon_0 - \delta\varepsilon_0 \cos(2\pi x/\Lambda)$, $\delta\varepsilon_0/\varepsilon_0 \ll 1$, resulting in the slight transverse modulation in the field, $\delta E/E_a = (\delta\varepsilon_0/\varepsilon_0)\cos(2\pi x/\Lambda)$, that plays the role of a small seed for the instability of interest.

The long-scale longitudinal structure can be described by the parabolic wave equation for the time-envelope of the average field

$$\frac{1}{S}\frac{\partial}{\partial z}\left(S\frac{\partial E_a}{\partial z}\right) + \frac{2i\omega}{c^2}\frac{\partial E_a}{\partial t} + \varepsilon_{eff} E_a = 0, \qquad (2)$$

where $\varepsilon_{eff}(z,t) = (1/\Lambda E_a)\int_0^\Lambda \varepsilon(x) E(x)\,dx$ is the effective permittivity determined by the small-scale transverse distributions $E(x)$ and $\varepsilon(x)$. The conventional Laplace operator is written in the wave equation (2) in the form $S^{-1}\partial/\partial z(S\partial/\partial z)$ allowing for the 1D modeling the longitudinal changes of the average field in the near-axial region of the paraxial beam. The function $S(z)$ is actually the cross-section area of some given ray tube, characterizing the convergence or divergence of the beam and determining according to Eq.(2) the behavior of the function $E_a(z)$. In what follows we set $S(z) = 1/[1+(z/l_F)^2]$ (see also [25]) as modeling the ray convergence in some axial-symmetric Gaussian beam with focal plane $z=0$ and the effective length of the focal region $l_F \gg k^{-1}$; the waist of this beam $a_F = \sqrt{l_F/k}$ and the characteristic convergence angle $\vartheta = 1/\sqrt{kl_F} \ll 1$ where $k = 2\pi/\lambda = (\omega/c)\sqrt{\varepsilon_0}$, $\lambda$ is the laser wavelength in the unperturbed medium. In the absence of ionization, the time-dependent solution of Eq. (2) is supposed to be the Gaussian pulse propagating in the $+z$ direction:

$$E_a^{(0)} = E_F \left(1 + \frac{z^2}{l_F^2}\right)^{-1/2} \exp\left(ikz - \frac{(t-z/c)^2}{\tau_p^2}\right). \quad (3)$$

Here $E_F$ is the maximal unperturbed field amplitude of the incident pulse at the focus, $\tau_p$ is the pulse width. The boundaries of the calculation region for Eq. (1) ($-L_1 < z < +L_2$) include the focal region and are assumed to be far enough away from it ($L_{1,2} > l_F$), so that the field amplitude at these boundaries is appreciably attenuated and the ionization is practically absent. The boundary conditions for Eq. (2), in this case, can be written in the paraxial approximation as the radiation conditions for the transmitted ($\sim \exp(ikz)$) and reflected ($\sim \exp(-ikz)$) waves in the unperturbed medium:

$$\text{at } z = -L_1: \quad \frac{\partial E_a}{\partial z} + ikE_a = 2ikE_a^{(i)}(-L_1, t), \quad (4)$$

$$\text{at } z = L_2: \quad \frac{\partial E_a}{\partial z} - ikE_a = 0. \quad (5)$$

With the most extensively studied nanostructuring conditions in mind (fused silica ionized by 800 nm 100 fs laser pulse), the electron density rate equation taking into account the multiphoton and electron impact ionization, recombination and diffusion has the form

$$\frac{\partial N}{\partial t} = W_{phi} + W_a - \frac{1}{\tau_r} N + D\left(\frac{\partial^2 N}{\partial x^2} + \frac{\partial^2 N}{\partial z^2}\right). \quad (6)$$

Here, $W_{phi}$ and $W_a$ are the rates of the multiphoton (6-photons for the fused silica) and avalanche (electron-impact-induced) ionization, respectively, determined by the known expressions [14,26,27]

$$W_{phi} = \sigma_6 I^6 (N_m - N), \quad \sigma_6 = 10^{-69} (\text{W}^{-6}\text{s}^{-1}\text{cm}^{12}), \quad (7)$$

$$W_a = \frac{8\pi \nu e^2 IN}{3\sqrt{\varepsilon_0} mcU(\omega^2 + \nu^2)} \cdot \frac{N_m - N}{N_m}, \quad (8)$$

$I = c\sqrt{\varepsilon_0}|E_x|^2/(8\pi)$ is the local intensity, $\varepsilon_0 = 2.1$, $N_m = 2.1 \times 10^{22} \text{cm}^{-3}$ is the atom density, $U = 9$ eV is the bandgap for the fused silica, $\tau_r = 150$ fs [28] is the characteristic time of electron recombination, $D$ is the ambipolar diffusion coefficient; its conjectural value (roughly estimated for high electron temperatures by far extrapolation of the results [29]) lays probably in the range between 10 and 100 $\text{cm}^2/\text{s}$.

The equation system (1), (2), and (6) was solved numerically with the radiation conditions (4) and (5) for the averaged field and the conditions of periodicity (with the period $\Lambda$ in the $x$-coordinate) for the field and density. The initial conditions corresponded to the absence of the field and ionization in the calculation region at large negative time ($N \to 0$ and $E_a \to 0$ $t \to -\infty$). The calculations were carried out at the above parameter values for the maximal unperturbed intensity in the focus $I_{max} = c\sqrt{\varepsilon_0} |E_F|^2 /(8\pi)$ lying in the range from $2.5 \times 10^{13}$ to $7 \times 10^{15}$ W/cm$^2$, the transverse modulation period $\Lambda = \lambda/3 \approx 190$ nm, the focal length $l_F = 35/k$ (the beam convergence angle $\vartheta = 10°$), the seed deviation of the medium permittivity $\delta \varepsilon_0 = 10^{-3}$; the electron collision frequency accordingly to Ref. [14,26] was taken $\nu = 0.15\omega$. Along with the functions $N(x,z,t)$ and $E(x,z,t)$, we also calculated the spatial distribution of the local energy deposition density in the medium $w(x,z,t)$ by the given time $t$ (see also [3,30]):

$$w = \int_{-\infty}^{t} \left[ \nu \frac{N}{N_c} |E|^2 + \left( 8\pi U + \frac{1}{2} \frac{|E|^2}{N_c} \right) s \frac{\partial N}{\partial t} \right] dt . \qquad (9)$$

Here, $s = 1$ at $\partial N/\partial t > 0$ and $s = 0$ at $\partial N/\partial t < 0$; the first term under integral describes the electron collision losses, the second one takes into account the energy expenditure for the interband transition (overcoming the bandgap $U$) and energy transfer to the newly born free electrons. This function is an important characteristic determining the rate and space profiles of the medium heating and therefore the course of the thermo-mechanical and chemical processes and accumulation effects [21,31] responsible for the volume nanograting formation in the medium by the repeated pulses. The high contrast nanograting can be formed evidently if by the end of the pulse the sufficient energy deposition $w(x)$ is localized in a comparatively narrow region in each period $\Lambda$ and there is a large enough difference between its maximal and minimal values.

The results of calculation are shown in Figs. 1-4. The typical scenarios of the plasma and field evolution within one spatial period ($-\Lambda/2 < x < +\Lambda/2$) are shown for three laser intensities in Figs. 1 and 2. The resulting spatial distributions of the energy deposition $w(x,z)$ at the end of the laser pulses for the same intensities are shown in Fig. 3. The curves in Fig. 4 show dependences of the maximum ($w_{max}$) and minimum ($w_{min}$) of the function $w(x)$ on the period at the point $z$ where the maximum is the largest. Along with them, there is shown in the same figure the intensity dependence of the maximum value $N_{max}/\varepsilon_0 N_c = 1 - (\text{Re}\,\varepsilon)_{min}/\varepsilon_0$ achieved during the breakdown pulse.

The typical discharge evolution bears the character of the breakdown wave propagating in the direction opposite to the incident laser beam (see Fig 1 and . Its parameters and transverse structure depend strongly on the incident pulse intensity and are closely connected with whether the plasma density passes in the breakdown process the plasma resonance point $\text{Re}\,\varepsilon = 0$, $N = \varepsilon_0 N_c$. As it is seen from Fig. 4, this is the case in some intensity range $I_1 < I_{max} < I_2$ (from $6.4 \times 10^{13}$ to $1.2 \times 10^{15}$ W/cm$^2$). In this range, the maximal plasma density $N_{max} > \varepsilon N_c$, the unstable perturbations evolve in the self-sharpening regime due to the field enhancement near plasma resonance ($E \sim 1/\varepsilon$), and are far ahead of the homogeneous background growth, resulting in the formation of a high contrast periodic plasma structure with rather narrow zones of the overcritical density and strong electric field (Fig. 1b, 2b). What is of prime importance, is that the energy deposition density achieves here the largest values ($w_{max} \approx$ 13-14 kJ) and is localized on each transverse period in rather narrow region ($\Delta x \approx \Lambda/4 \approx 45$ nm). Outside this intensity range, the maximal plasma density $N_{max} < \varepsilon_0 N_c$ and the resonance does not appear. At $I_{max} < I_1$, the growth rate of the instability is too small, the periodic perturbations of the field and density at the given small seed modulation ($\delta\varepsilon_0 = 10^{-3}$) have no time to become appreciable, the instability does not achieve its nonlinear stage, and the contrast transverse structure is not generated (Fig. 1a, 2a, 3a). At $I_{max} > I_2$, the instability evolves but at more slow rate (for lack of the resonance sharpening effect), resulting in the decrease of the maximal energy deposition and the contrast of the formed periodic structure. The existence of some optimal intensity range for the grating-like structure formation agrees with the experiments [8].

    To conclude, we have studied the dynamics and structure of the laser-pulse-produced discharge affected by the small-scale (plasma-resonance-induced) ionization instability within the volume of the transparent dielectric (fused silica). The instability originating from small seed periodic perturbations against the background of quasi-homogeneous discharge results at its nonlinear stage in the deep subwavelength modulation of the discharge parameters along the incident pulse polarization. The simplified model used has allowed us to take into account both the long-scale field variations (caused by the beam focusing, phase and group delays, and backward reflection) and the small-scale ones (quasi-static field enhancement in the plasma resonance region). The rate equation for the plasma density has allowed for the photoionization, avalanche (electron-impact induced) ionization, and the ambipolar diffusion. Based on the fulfilled numerical calculations we have found the laser pulse intensity range where the instability considered leads to the formation of the contrast periodic structure containing the number of the narrow zones with overcritical plasma density and enhanced energy deposition. The latter allows us to consider it as underlying the nanograting formation observed experimentally in the transparent dielectric irradiated by the series of repeated fs pulses.

This work was supported by the Russian Science Foundation (Grant No. 17-12-01574).

1. V.B. Gildenburg and A.V. Kim, Sov. Phys. JETP **47**, 72 (1978).
2. M. Lontano, G. Lampis, A.V. Kim, and A.M. Sergeev, Phys. Scr. **T63**, 141 (1996).
3. V.B. Gildenburg, A.G. Litvak, and N.A. Zharova, Phys. Rev. Lett. **78**, 2968 (1997).
4. T.M. Antonsen Jr and Z. Bian, Phys. Rev. Lett. **82**, 3617 (1999).
5. N.V. Vvedenskii and V.B. Gildenburg, JETP Lett. **76**, 380 (2002).
6. E.S. Efimenko, A.V. Kim, and M. Quiroga-Teixeiro, Phys. Rev. Lett. **102**, 015002 (2009).
7. Y. Shimotsuma, P.G. Kazansky, J. Qiu, and K. Hirao, Phys. Rev. Lett. **91**, 247405 (2003).
8. C. Hnatovsky, R.S. Taylor, P.P. Rajeev, E. Simova, V.R. Bhardwaj, D.M. Rayner, and P.B. Corkum, Appl. Phys. Lett. **87**, 014104 (2005).
9. V.R. Bhardwaj, E. Simova, P.P. Rajeev, C. Hnatovsky, R.S. Taylor, D.M. Rayner, and P.B. Corkum, Phys. Rev. Lett. **96**, 057404 (2006).
10. R. Taylor, C. Hnatovsky, and E. Simova, Las. Photon. Rev. **2**, 26 (2008).
11. M. Huang, F. Zhao, Y. Cheng, N. Xu, and Z. Xu, ACS Nano **3**, 4062 (2009).
12. C. Mauclair, M. Zamfirescu, J.-P. Colombier, G. Cheng, K. Mishchik, E. Audouard, and R. Stoian, Opt. Expr. **20**, 12997 (2012).
13. M. Beresna, M. Gecevičius, P.G. Kazansky, T. Taylor, and A.V. Kavokin, Appl. Phys. Lett. **101**, 053120 (2012).
14. N.M. Bulgakova, V.P. Zhukov, and Y.P. Meshcheryakov, Appl. Phys. B**113**, 437 (2013).
15. S. Richter, M. Heinrich, F. Zimmermann, C. Vetter, A. Tünnermann, and S. Nolte, "Nanogratings in fused silica: Structure, formation and applications," in Progress in Nonlinear Nano-Optics, (Springer, 2015), pp. 49–71.
16. R. Buschlinger, S. Nolte, and U. Peschel, Phys. Rev. B **89**, 184306 (2014).
17. Y. Liao, J. Ni, L. Qiao, M. Huang, Y. Bellouard, K. Sugioka, and Y. Cheng, Optica **2**, 329 (2015).
18. Y. Liao, W. Pan, Y. Cui, L. Qiao, Y. Bellouard, K. Sugioka, and Y. Cheng, Opt. Lett. **40**, 3623 (2015).
19. A. Rudenko, J.-P. Colombier, and T.E. Itina, Phys. Rev. B **93**, 075427 (2016).
20. A. Rudenko, J.-P. Colombier, S. Höhm, A. Rosenfeld, J. Krüger, J. Bonse, and T.E. Itina, Sci. Rep. **7**, 12306 (2017).
21. A. Rudenko, J.-P. Colombier, and T.E. Itina, Phys. Chem. Chem. Phys. **20**, 5887 (2018).
22. V.B. Gildenburg and I.A. Pavlichenko, Phys. Plas. **23**, 084502 (2016).
23. V.B. Gildenburg and I.A. Pavlichenko, Phys. Plas. **24**, 122306 (2017).
24. L. Wang, Q.-D. Chen, X.-W. Cao, R. Buividas, X. Wang, S. Juodkazis, and H.-B. Sun, Light. Sci. Appl. **6**, e17112 (2017).
25. V.E. Semenov, E.I. Rakova, V.P. Tarakanov, M.Yu. Glyavin, and G.S. Nusinovich, Phys. Plas. **22**, 092308 (2015).
26. A. Couairon, L. Sudrie, M. Franco, B. Prade, and A. Mysyrowicz, Phys. Rev. B **71**, 125435 (2005).
27. K.I. Popov, C. McElcheran, K. Briggs, S. Mack, and L. Ramunno, Opt. Expr. **19**, 271 (2011).
28. P. Audebert, Ph. Daguzan, A. Dos Santos, J.C. Gauthier, J.P. Geindre, S. Guizard, G. Hamoniaux, K. Krastev, P. Martin, G. Petite, and A. Antonetti, Phys. Rev. Lett. **73**, 1990 (1994).
29. R.C. Hughes, Appl. Phys. Lett. **26**, 436 (1975).
30. V.B. Gildenburg, A.V. Kim, V.A. Krupnov, V.E. Semenov, A.M. Sergeev, and N.A. Zharova, IEEE Trans. Plasma Sci. **21**, 34 (1993).
31. N.M. Bulgakova, V.P. Zhukov, S.V. Sonina, and Yu.P. Meshcheryakov, J. Appl. Phys. **118**, 233108 (2015).

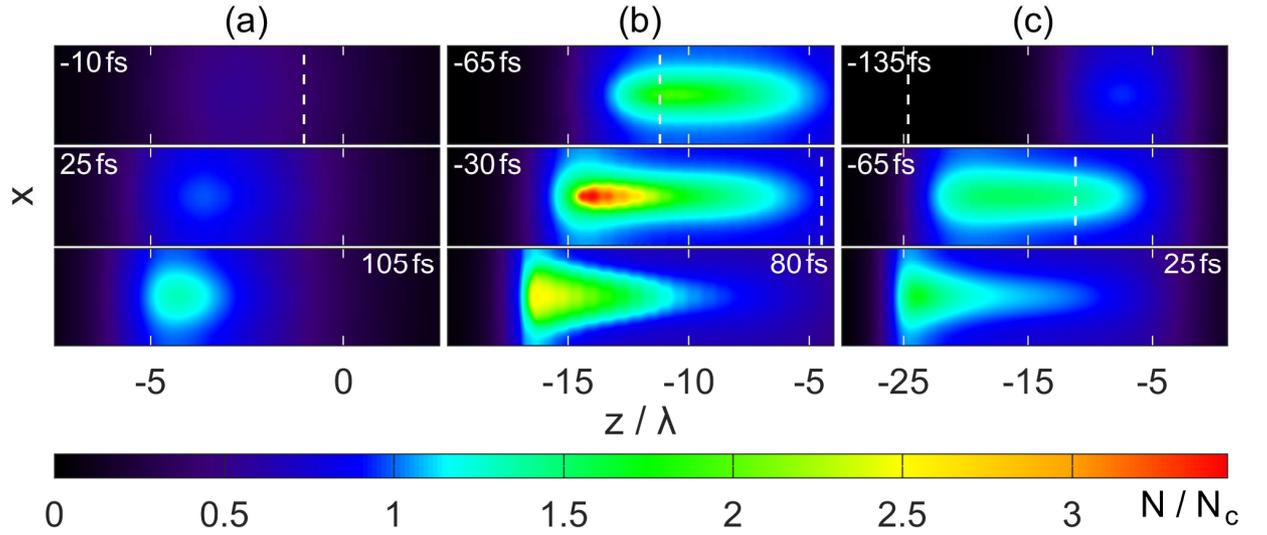

Fig.1. Spatiotemporal evolution of the focused-laser-pulse-produced discharge affected by the plasma resonance ionization instability in the fused silica for three different maximal intensities of the unperturbed laser pulse: (a) $5\times10^{13}$, (b) $4\times10^{14}$, and (c) $1.5\times10^{15}$ W/cm$^2$. The laser pulse propagates in $z$-direction from left to right. The focal plane coordinate $z=0$. The snapshots show the spatial plasma density distributions $N(x,z)$ on one spatial period ($\Lambda = 190$ nm) of its transverse ($x$-periodic) structure at different time instants (increasing in each column from top to bottom). Laser pulse parameters: wavelength 800 nm, width 100 fs, convergence angle 10°. Dashed vertical lines show the position of the pulse maximum in the absence of plasma.

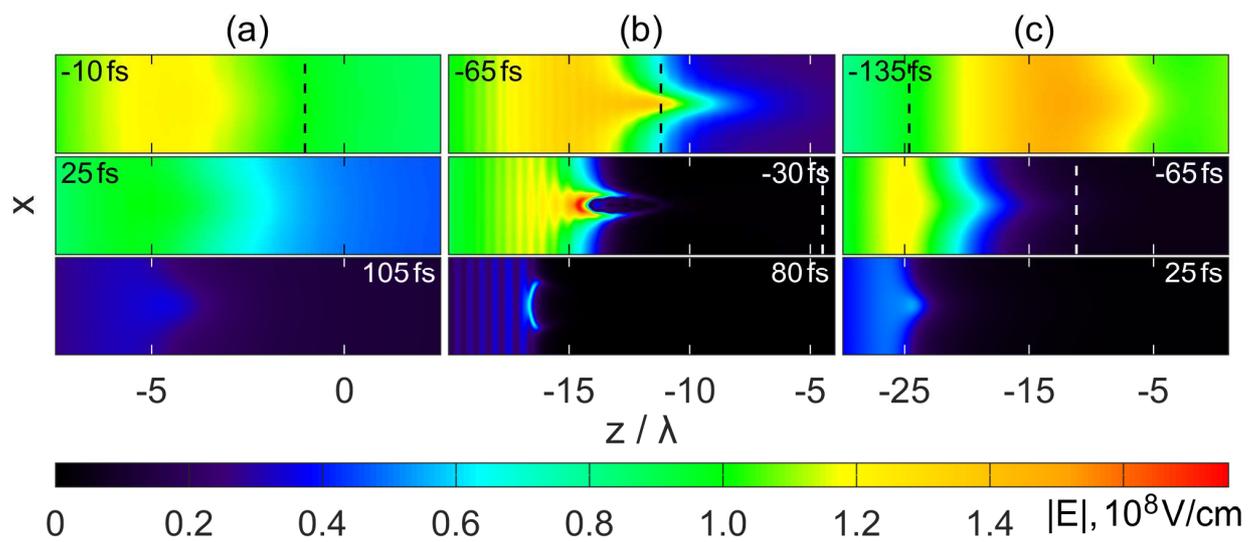

Fig.2. Snapshots of the electric field amplitude $|E|(x,z,t)$ at the same time instants and for the same laser intensities and pulse parameters as in Fig. 1.

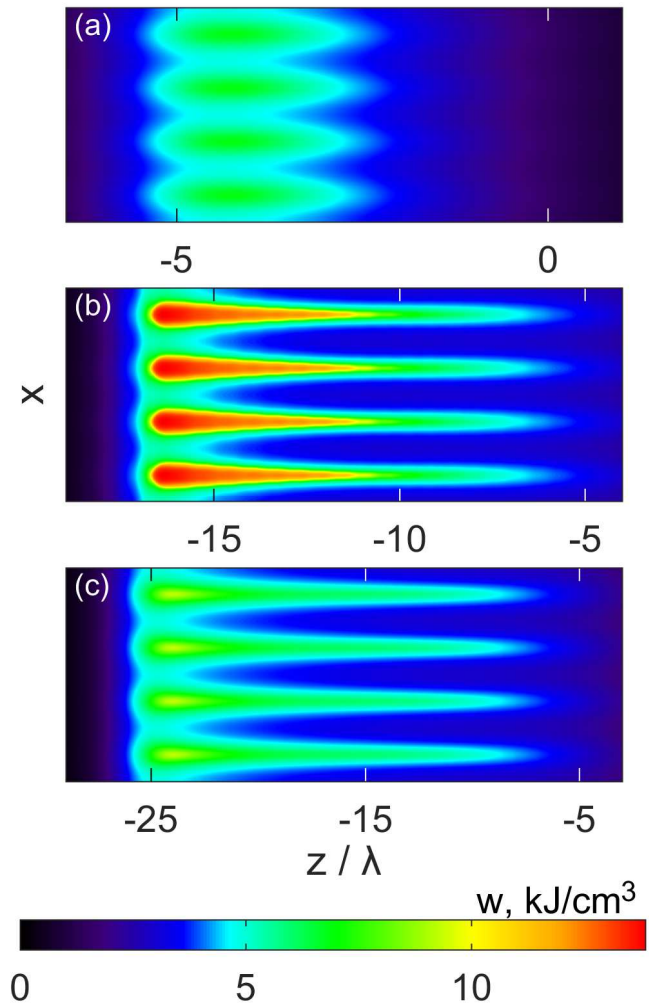

Fig. 3. Spatial distributions of the resulting energy deposition density $w(x,z)$ at the end of the breakdown pulse in the fused silica under the same conditions as in Fig. 1 (to illustrate with the most clearness the "grating character" of the calculated structures we show in this figure several periods in $x$-direction).

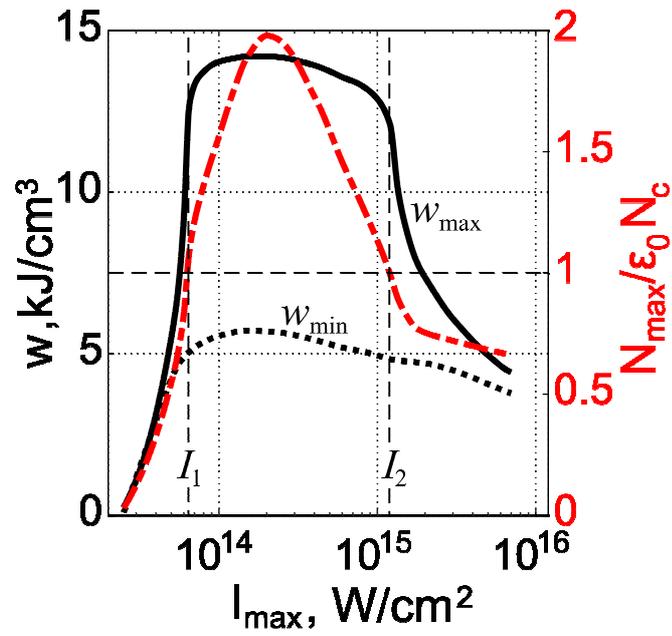

Fig.4. Laser intensity dependences of the maximal ($w_{max}$) and minimal ($w_{min}$) energy deposition densities within each spatial period (the solid and dotted line, respectively) and of the maximal plasma density (normalized to the value $\varepsilon_0 N_c$) at the same laser parameters as in Fig.1.